\documentclass{article}
\usepackage{glas}
\usepackage{epsf}
\epsfverbosetrue
\def\mb#1{\mbox{\scriptsize {#1}}}
\def\mt#1{\mbox{\tiny {#1}}}
\def\Journal#1#2#3#4{{#1} {\bf #2}, (#4) #3}
\def\PRD{{\em Phys. Rev.} D}
\def\PRL{\em Phys. Rev. Lett.}
\def\PLB{{\em Phys. Lett.}  B}
\def\NPB{{\em Nucl. Phys.} B}
\def\CPC{\em Comp. Phys. Comm.}
\def\oos{${\cal O}(\alpha \alpha_s^2)$}
\begin{document}
\begin{titlepage}{GLAS-PPE/98--05}{October 1998}
\title{Jets and Prompt Photons in Photoproduction at ZEUS}
\author{L.\ E.\ Sinclair}
\collaboration{for the ZEUS Collaboration}
\begin{abstract}
In the ZEUS experiment at HERA, photoproduction processes have been 
studied for photon-proton
centre-of-mass energies in the range $100 < W_{\gamma p} < 300$~GeV and
jet transverse energies extending to $E_T^{\mt{jet}} \sim 70$~GeV.
The data contribute to our understanding of QCD dynamics, and also
provide new constraints on the photon's parton density.
\end{abstract}
\vfill
\conference{talk given at the XXIX International Conference on
High Energy Physics (ICHEP '98) \\
Vancouver, Canada.  August, 1998.}
\end{titlepage}
\section{Introduction}

There has been tremendous progress in the understanding of photon induced 
reactions since the start of HERA operation.
Up to the end of 1994, HERA had delivered a total of 7~pb$^{-1}$ of
$ep$ data.  The analysis of this data was fruitful.  ZEUS was able to 
clearly establish evidence for both the
direct and the resolved classes of process.  
These are illustrated at leading order (LO) in 
Figure's~1(a) and (b).
\begin{figure}[htb]
\center
\epsfxsize=2.5cm
\leavevmode
\epsfbox{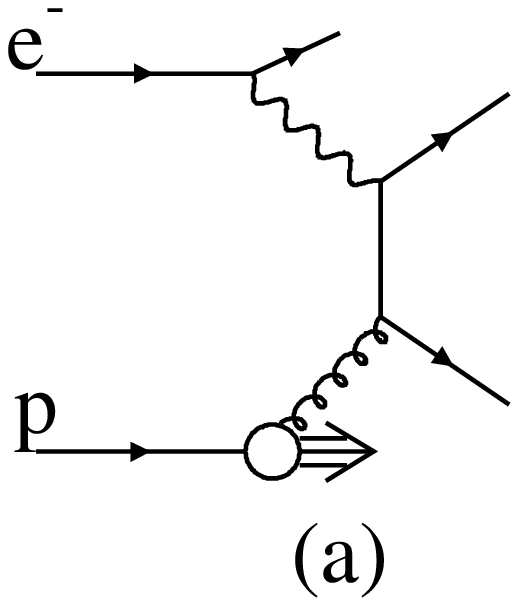}
\hspace{.4cm}
\epsfxsize=2.5cm
\leavevmode
\epsfbox{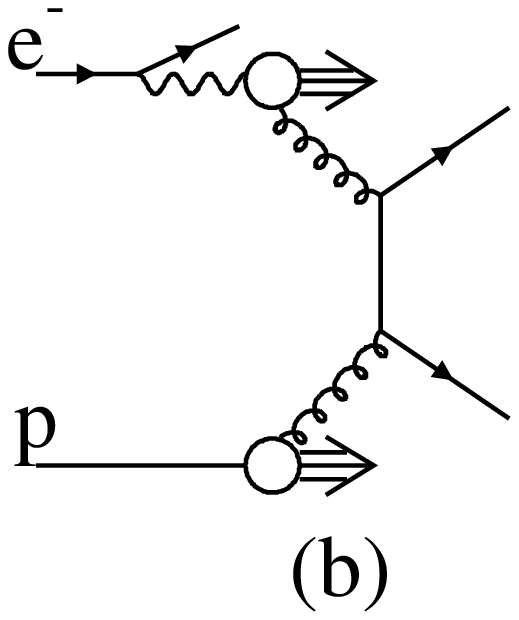}
\hspace{.4cm}
\epsfxsize=2.5cm
\leavevmode
\epsfbox{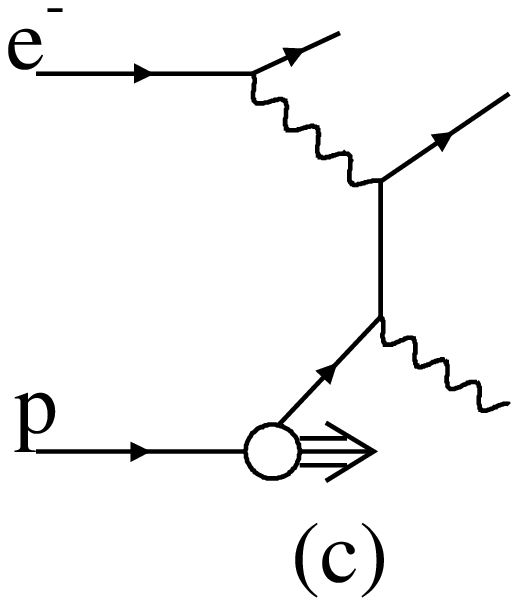}
\hspace{.4cm}
\caption{Examples of leading order diagrams for (a) direct photoproduction, 
(b) resolved 
photoproduction and (c) direct prompt photon photoproduction.}
\label{fig:diagrams}
\end{figure}

An observable quantity was defined which extended the distinction between
direct and resolved processes to all orders of perturbation theory,
\begin{math}
x_{\gamma}^{\mb{OBS}} = 
               \sum_{\mb{jets}} E_T^{\mb{jet}} e^{-\eta^{\mt{jet}}}
                    / 2E_{\gamma}
\end{math}
where the sum runs over the two highest transverse energy jets in the
final state.  Direct processes contribute primarily at high 
$x_{\gamma}^{\mb{OBS}}$ and resolved processes contribute at low
$x_{\gamma}^{\mb{OBS}}$.
However the comparison with theory has been hampered by the 
need to introduce 
models such as secondary interactions between the photon and proton
constituents in order to understand the resolved processes at 
$E_T^{\mb{jet}} \sim 6$~GeV.

\begin{figure}[htb]
\centering
\leavevmode
\epsfxsize=8.cm
\epsfbox{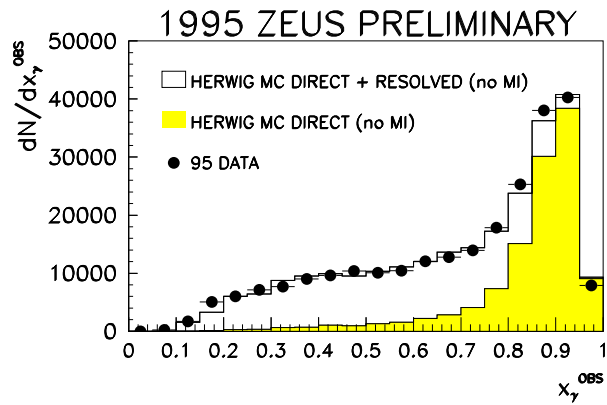}
\caption{Distribution of $x_{\gamma}^{\mt{OBS}}$.  The black dots show 
the uncorrected data.  The histogram shows the HERWIG prediction with
the contribution due to LO direct processes shaded.  Multiple interactions
have not been included in the simulation.}
\label{fig:xgamma}
\end{figure}

By the end of 1997 running HERA had delivered an order of magnitude
more data, $\sim 70$pb$^{-1}$.
This large data-set allows the exploration of photoproduction up
to much higher jet transverse energies where a strong confrontation
with the predictions of perturbative QCD (pQCD) may 
be made.
This is illustrated in Figure~\ref{fig:xgamma}~\cite{ichep_810}.  Here
the $x_\gamma^{\mb{OBS}}$ distribution is shown for photoproduction
events where at least two jets are produced, one with
$E_T^{\mb{jet}} > 14$~GeV and the other with 
$E_T^{\mb{jet}} > 11$~GeV where the jet pseudorapidities satisfy
$-1 < \eta^{\mb{jet}} < 2$.  The results are compared with the 
Monte Carlo predictions from HERWIG~\cite{HERWIG}, which 
implements the leading order matrix elements together with a parton 
shower.  A good description of the data is provided by HERWIG
across the entire $x_\gamma^{\mb{OBS}}$ range, without having to resort
to additional parameters.  This facilitates the interpretation of the
data within pQCD.

This high accumulated luminosity also gives access to such infrequent
processes as prompt photon production, illustrated in
Figure~\ref{fig:diagrams}(c),
and multijet production.
In this talk results are presented on four topics; 
single and dijet inclusive cross sections~\cite{ichep_810,ichep_812},
prompt photon production~\cite{ichep_815},
the evolution of the photon's structure with virtuality~\cite{ichep_816} 
and two and three-jet angular distributions in high mass 
events~\cite{ichep_805,ichep_800}.

\section{Single and Dijet Inclusive Cross sections}

In hadron-hadron collisions, jets of final-state hadrons are commonly
determined by cone algorithms which seek to maximize the transverse
energy produced inside a cone of fixed radius in pseudorapidity and
azimuth.  However the confrontation of data and theory using such 
algorithms is compromised by the need to introduce an ad-hoc parameter
$R_{\mb{SEP}}$ into the theory, in order to simulate merging 
effects~\cite{RSEP}.
A clustering algorithm which combines particles with small relative
transverse momenta, $k_T$, into jets has been developed for use in
hadron-hadron collisions~\cite{Mike_93,Ellis_93}.  With this
algorithm it is not necessary to introduce additional parameters 
in order to compare data and theory.  In the results presented here the
$k_T$ clustering algorithm has been used.

\begin{figure}[htb]
\centering
\leavevmode
\epsfxsize=5.7cm
\epsfbox{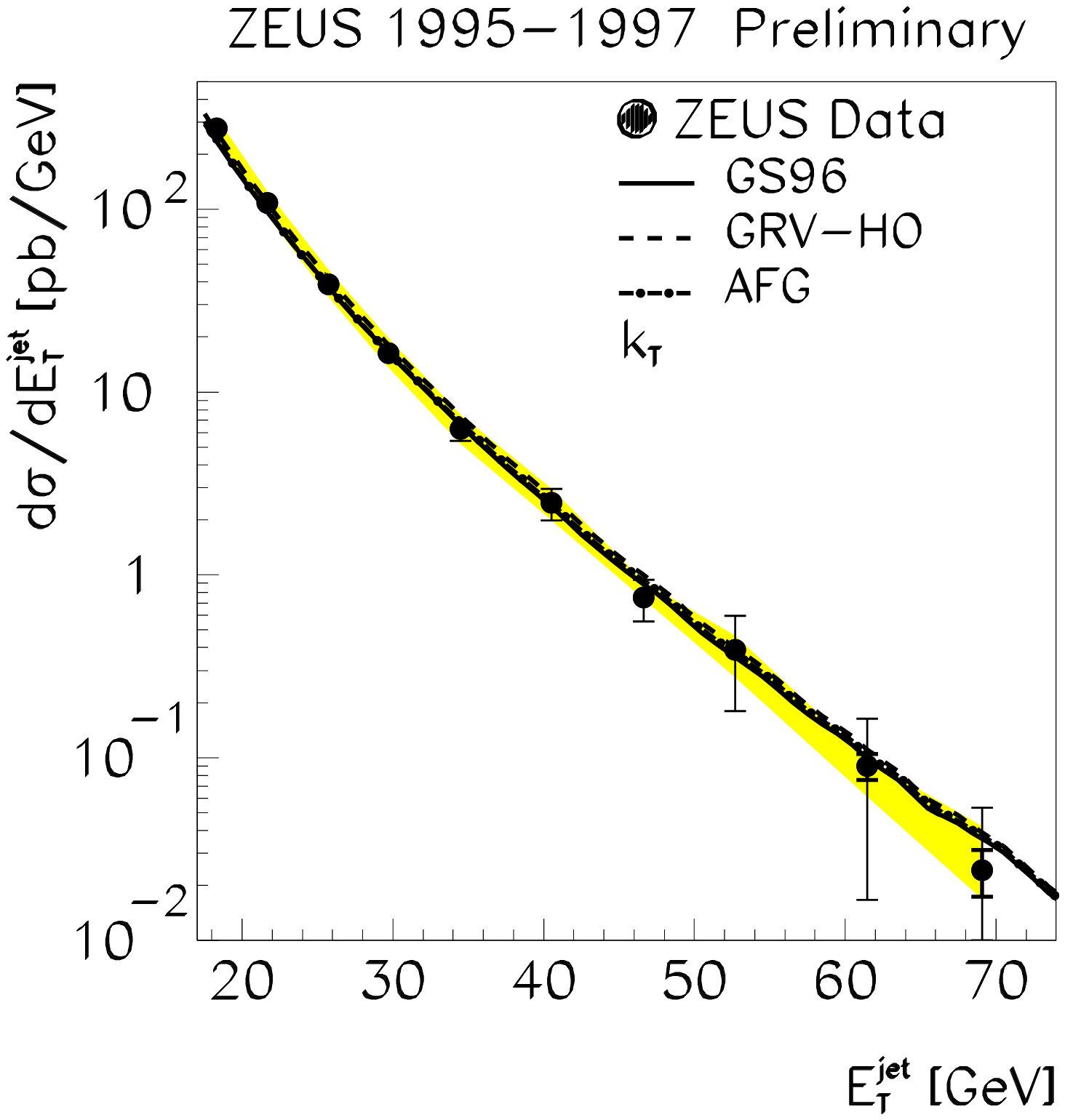}\epsfxsize=5.7cm\epsfbox{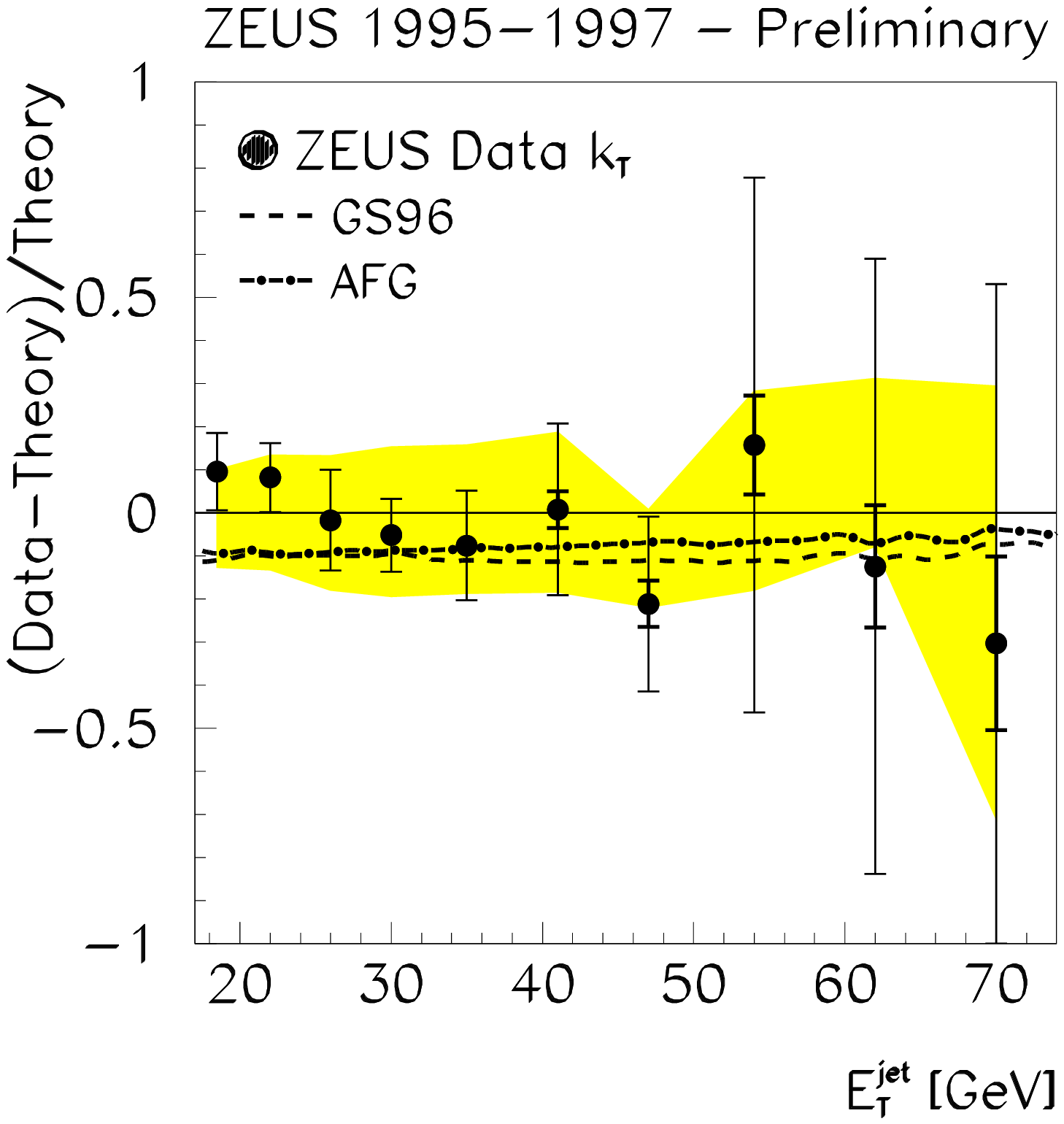}
\caption{Left: Inclusive jet production cross section, 
$d\sigma/dE_T^{\mt{JET}}$.  Right: Fractional difference between the
measured cross section and the pQCD prediction using the GRV~HO 
parametrization of the photon parton densities.  The dots show the data.
The inner error bars show the statistical error.  The outer error bars
show the quadratic sum of the statistical and systematic uncertainties
with the exception of the absolute energy scale uncertainty which
is shown separately as a shaded band.}
\label{fig:incjets}
\end{figure}

Figure~\ref{fig:incjets} presents the inclusive jet cross 
section, $d\sigma/dE_T^{\mb{jet}}$, for jets with
$-0.75 < \eta^{\mb {jet}} < 2.5$ and for centre-of-mass (CM)
energies in the range $134 < W_{\gamma p} < 277$~GeV~\cite{ichep_812}.
The cross section falls by 4 orders of magnitude over the $E_T^{\mb{jet}}$
range, extending to $\sim 70$~GeV.  This behaviour is impressively
reproduced in the theory.  The three curves show \oos\ pQCD
calculations for different parametrizations of the parton densities 
of the photon.  Within the current measurement uncertainties all are
consistent with the data.  This is more clear in the plot on the right
which
shows the relative difference with respect to the calculation using
the GRV~HO parametrization.

\begin{figure}[htb]
\centering
\leavevmode
\epsfxsize=10.cm
\epsfbox{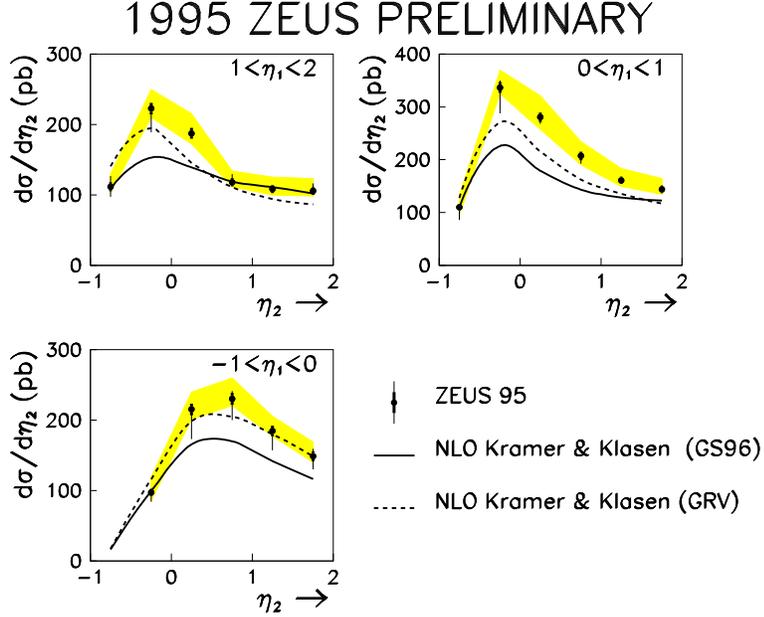}
\caption{Dijet inclusive cross section, $d\sigma/d\eta_2^{\mt{jet}}$.
Error bars are as described for Figure~\ref{fig:incjets}.
NLO calculations for two 
parametrizations of the photon parton distributions are shown by the 
solid and dashed curves.}
\label{fig:dijets}
\end{figure}

The dijet inclusive photoproduction cross section as a function of
the pseudorapidity of one of the jets,
$d\sigma/d\eta_2$, is shown in Figure~\ref{fig:dijets}, in three
bins of the pseudorapidity of the other jet, 
$\eta_1$~\cite{ichep_810}.
Here the events have been required to have at least one jet
with $E_T^{\mb{jet}} > 14$~GeV and at least one other with
$E_T^{\mb{jet}} > 11$~GeV.  The photon-proton CM
energies have been limited to the range 
$212 < W_{\gamma p} < 277$~GeV,
in order to enhance sensitivity to the photon's parton 
densities.  Indeed, by comparison with the next-to-leading order
(NLO) calculations we find that one parton density set
(GRV) is favoured over another (GS96).  It is worth emphasizing
that GS96 is a modern set, consistent
with the world data on $F_2^\gamma$.  The ability of the 
data to disfavour this, indicates that ZEUS is
contributing photoproduction information in a heretofore unexplored 
regime.

\section{Prompt Photon Production}

A clean laboratory for testing QCD predictions is provided by 
prompt photon production,
Figure~\ref{fig:diagrams}(c).  
An inclusive prompt photon measurement, tagging only the final-state
hard photon, may be directly compared with pQCD calculations,
with neither concern for jet definition matching between theory and 
experiment, nor for hadronization uncertainties.

\begin{figure}[h!]
\centering
\leavevmode
\begin{picture}(100,2)(0,0)
\put(0.,2.){\sf{ZEUS 1996 - 1997 Preliminary}}
\end{picture}

\centering
\leavevmode
\epsfxsize=6.cm
\epsfbox{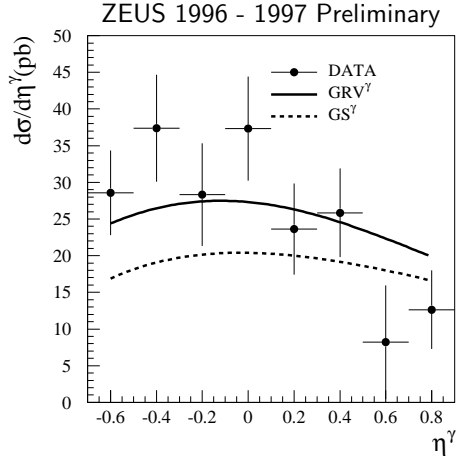}
\caption{Inclusive prompt photon photoproduction cross section,
$d\sigma/d\eta^{\gamma}$.  The errors shown are statistical only.
The lines show QCD NLO predictions for two different
parametrizations of the photon parton densities.  A systematic
uncertainty of $\pm 15$~\% should be allowed.}
\label{fig:prompt}
\end{figure}

In Figure~\ref{fig:prompt} the inclusive prompt photon cross section
with respect to the photon's pseudorapidity, $d\sigma/d\eta^\gamma$,
is shown.  The events lie in the range 
$120 < W_{\gamma p} < 270$~GeV and have
at least one photon of transverse energy $5 < E_T^\gamma < 10$~GeV.
The cross section is compared with predictions for two different 
parametrizations of the photon's structure, GRV and GS.  Again we
find that there is a preference for the GRV parton densities over
the GS parton densities.

\section{Virtual Photon Structure}

In addition to asking about the parton density of the real photon,
one can ask how that structure will evolve with the negative
invariant mass squared of the photon, or the photon's virtuality,
$Q^2$.  One expects that as $Q^2$ increases, less time is available
for the photon to develop a complicated hadronic structure.

\begin{figure}[htb]
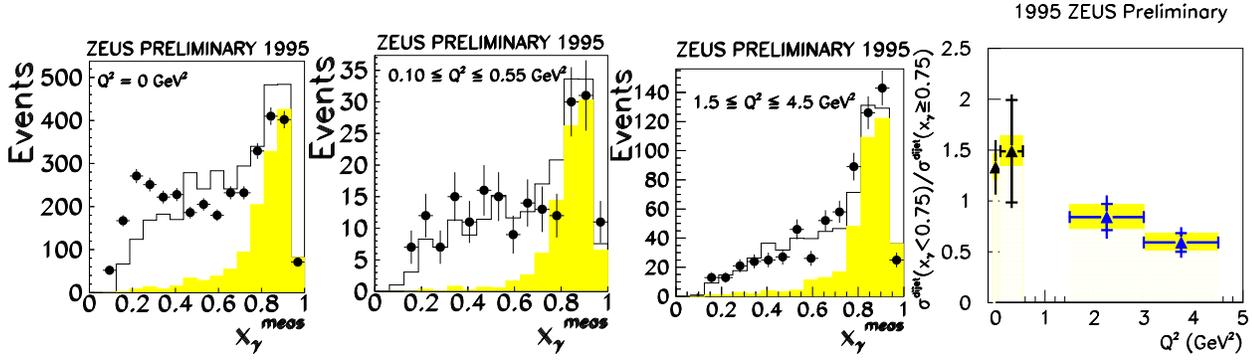

\epsfxsize=4.cm
\leavevmode
\epsfbox{xg_low.epsi}
\epsfxsize=4.cm
\leavevmode
\epsfbox{xg_medium.epsi}
\epsfxsize=4.cm
\leavevmode
\epsfbox{xg_high.epsi}
\epsfxsize=4.5cm
\leavevmode
\epsfbox{final_kt_ratio.epsi}
\caption{Uncorrected $x_{\gamma}^{meas}$ ($x_{\gamma}^{\mt{OBS}}$) 
distributions in three photon virtuality ranges where the error
bars show the statistical errors only.  The histogram
shows the HERWIG prediction with the LO direct contribution shaded.
Also, the ratio of the resolved to the direct dijet 
cross sections versus $Q^2$\label{fig:virtual} where the error bars are
as described for Figure~\ref{fig:incjets}.}
\end{figure}

In Figure~\ref{fig:virtual}, the uncorrected distribution of
$x_{\gamma}^{\mb{OBS}}$ is shown 
for events with at least two jets of
$E_T^{\mb{jet}} > 6.5$~GeV and $-1.125 < \eta^{jet} < 1.875$,
in three different ranges of $Q^2$.
These distributions indicate that the frequency of resolved processes 
(which populate low $x_{\gamma}^{\mb{OBS}}$) decreases with respect to the 
frequency of direct processes, as $Q^2$ increases.
To quantify this observation, the ratio of the dijet inclusive cross 
section for $x_\gamma^{\mb{OBS}} < 0.75$ to the dijet inclusive cross 
section for $x_\gamma^{\mb{OBS}} \ge 0.75$ has been measured as a 
function of $Q^2$.  This is also shown in Figure~\ref{fig:virtual}.
This measurement confirms the expectation that resolved photoproduction
processes are supressed as the photon's virtuality increases.

\section{High Mass Dijet and Three-jet Distributions}

\begin{figure}[htb]
\centering
\leavevmode
\epsfxsize=7.5cm
\epsfbox{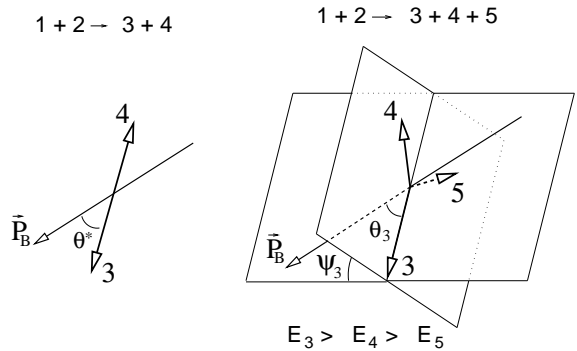}
\caption{Centre-of-mass frame diagrams of two and three-body scattering angles.  The beam
direction is indicated by $\vec{P}_{B}$.}
\label{fig:frames}
\end{figure}

As well as probing the partons within the photon, the
ZEUS photoproduction data can be used to learn something about 
the dynamics of
parton-parton scattering.  The definition of the scattering angles in 
two and three-jet events are
illustrated in Figure~\ref{fig:frames}.

\begin{figure}[h!]
\centering
\leavevmode
\epsfxsize=5.7cm
\epsfbox{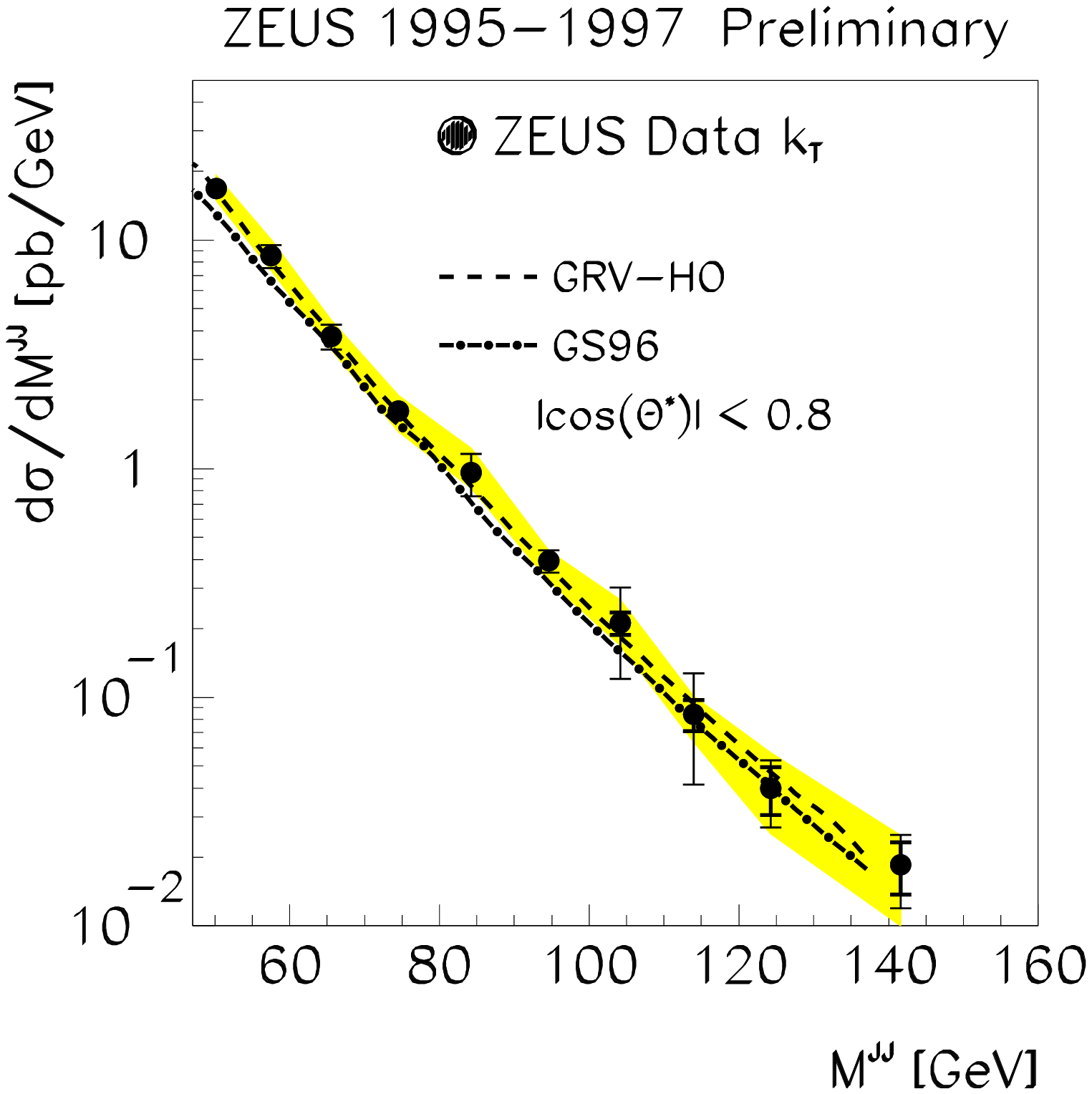}\epsfxsize=5.7cm\epsfbox{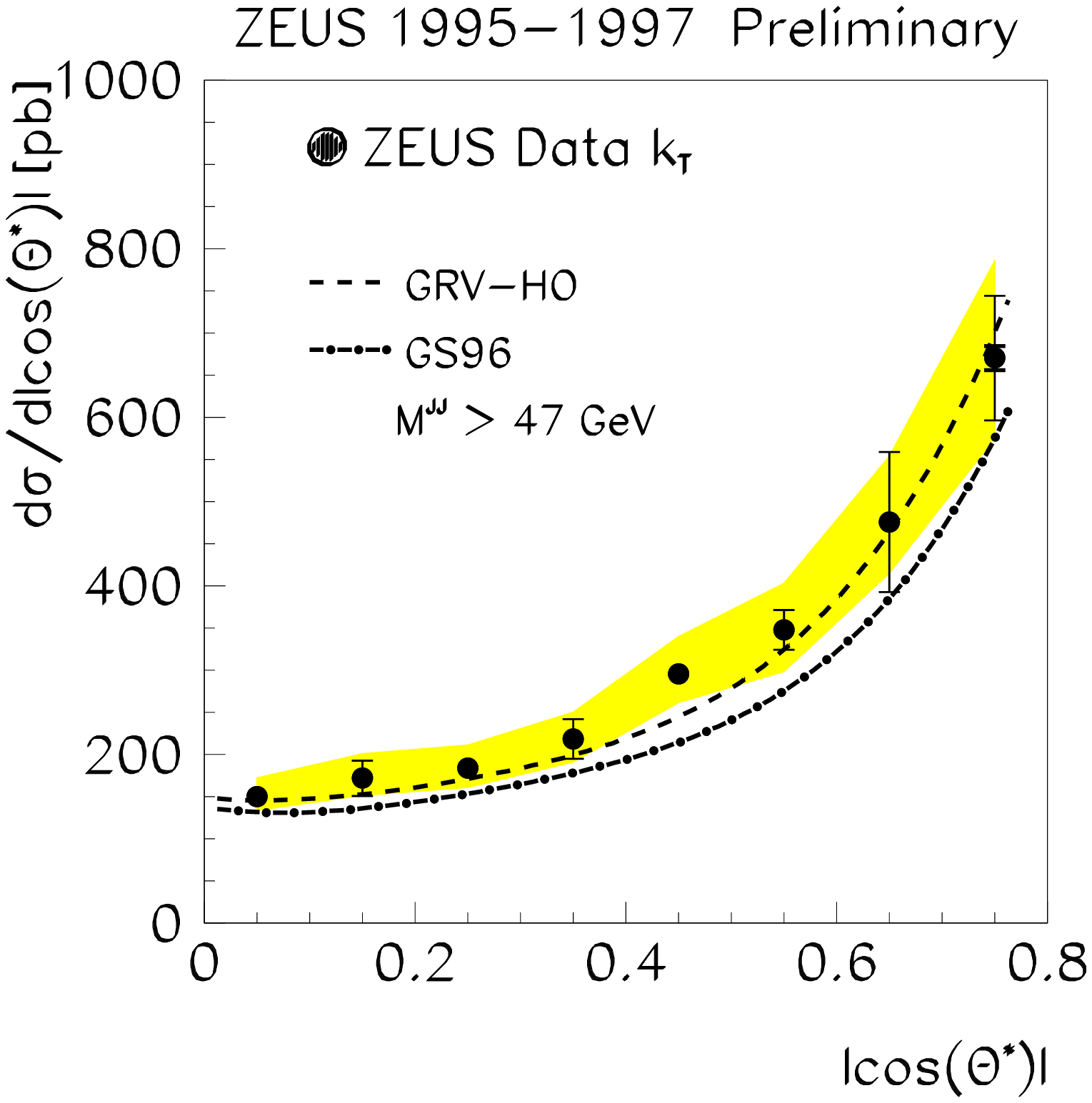}
\caption{Left: Differential dijet cross section, $d\sigma/dM^{\mb{JJ}}$.
Right: Differential dijet cross section, $d\sigma/d|\cos \theta^*|$.
The error bars are as described for Figure~\ref{fig:incjets}.  NLO 
QCD calculations for two parametrizations of the photon parton 
distributions are shown by the dashed and dot-dashed lines.}
\label{fig:dij_mjj}
\end{figure}

In
Figure~\ref{fig:dij_mjj} the dijet invariant mass distribution is shown for events
in which the scattering angle in the dijet rest frame, $\theta^*$, 
satisfies $|\cos \theta^*| < 0.8$~\cite{ichep_805}.
Resonances decaying to two jets would be expected to show an excess
above the pQCD predictions in a limited range of 
$M^{\mb{JJ}}$.  No evidence for this is seen.

The scattering angle in the dijet CM frame in photoproduction has been
shown to be sensitive to the spin of the exchanged 
parton~\cite{dijet_96}.
The $\cos \theta^*$ distribution is shown for the events with 
$M_{\mb{JJ}} > 47$~GeV on the right in
Figure~\ref{fig:dij_mjj}.
The NLO pQCD calculations are in good agreement with the data 
confirming that the dynamics of two-jet production are well understood.

\begin{figure}[htb]
\centering
\leavevmode
\epsfxsize=6.cm
\epsfbox{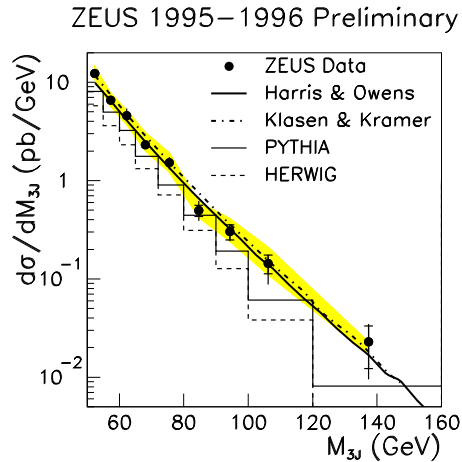}
\caption{The three-jet cross section $d\sigma/dM_{\mt{3J}}$
The error bars are as described for Figure~\ref{fig:incjets}.
\oos\ pQCD calculations
by two groups of authors are shown by thick solid and dot-dashed lines.
The thin solid and dashed histograms show the predictions from PYTHIA
and HERWIG.}
\label{fig:M3J}
\end{figure}

ZEUS has also measured the high-mass three-jet cross section,
$d\sigma/dM_{\mb{3J}}$, as shown in
Figure~\ref{fig:M3J}~\cite{ichep_800}.
The \oos\ pQCD calculations from two groups of authors 
(Harris \& Owens, Klasen
\& Kramer) provide a good description of the data, even though they
are leading order for this process.  Monte Carlo models also generate
three-jet events through the parton shower mechanism and both
PYTHIA~\cite{PYTHIA} and HERWIG reproduce the shape of the 
$M_{\mb{3J}}$ distribution.

\begin{figure}[htb]
\centering
\leavevmode
\epsfxsize=12.cm
\epsfbox{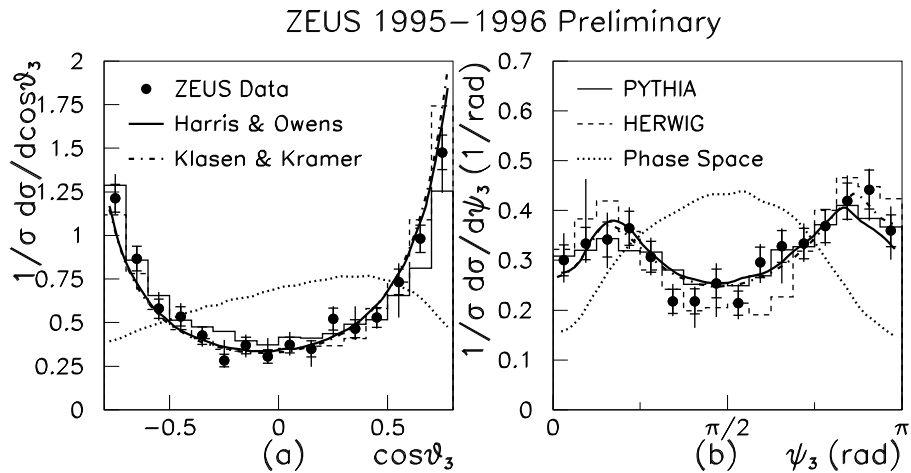}
\caption{Distribution of the angles $\cos \theta_3$ and $\psi_3$.  The
dotted curve shows the distribution for a constant matrix element.
Other details are as described for Figure~\ref{fig:M3J}.}
\label{fig:angles}
\end{figure}

For three-jet events there are two relevant scattering angles as
illustrated in Figure~\ref{fig:frames}.  The distributions of
$\cos \theta_3$ and $\psi_3$ are shown in
Figure~\ref{fig:angles}.
The $\cos \theta_3$ distribution resembles that of $\cos \theta^*$ and
exhibits forward and backward peaks.  It is well described in both
\oos\ pQCD calculations and parton shower models.  The $\psi_3$ 
distribution is peaked near 0 and $\pi$ indicating that the three-jet
plane tends to lie near the plane containing the highest energy jet
and the beam.  This is particularly evident if one considers the
$\psi_3$ distribution for three partons uniformly distributed in the
available phase space.  The phase space near $\psi_3 = 0$ and $\pi$
has been depleted by the $E_T^{\mb jet}$ cuts and by the jet-finding
algorithm.  The pQCD calculations describe perfectly the $\psi_3$
distribution.  It is remarkable that the parton shower models PYTHIA
and HERWIG are also able to reproduce the $\psi_3$ distribution.

\begin{figure}[htb]
\centering
\leavevmode
\epsfxsize=12.cm
\epsfbox{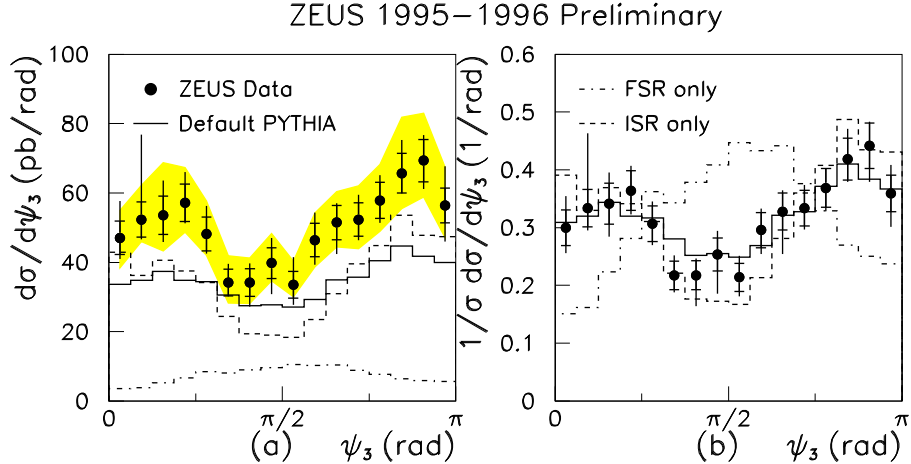}
\caption{(a) The cross section $d\sigma/d\psi_3$ and (b) the normalized 
distribution of $\psi_3$.  The error bars are as described for
Figure~\ref{fig:incjets}.  The solid line shows the default PYTHIA prediction,
the dashed line shows the prediction with FSR switched
off and the dot-dashed line shows the distribution with ISR switched off.}
\label{fig:psi_isr}
\end{figure}

Within the parton-shower model it is possible to determine the
contribution to three-jet production from initial-state radiation
(ISR) and final-state
radiation (FSR).  In
Figure~\ref{fig:psi_isr} the cross section $d\sigma/d\psi_3$ is shown,
along with the area-normalized $\psi_3$ distribution.
Here the PYTHIA prediction is shown for FSR only, 
ISR only, and default PYTHIA which
includes the interference of these two.
Clearly, ISR is predominantly responsible for
the three-jet production.

\begin{figure}[htb]
\centering
\leavevmode
\epsfxsize=12.cm
\epsfbox{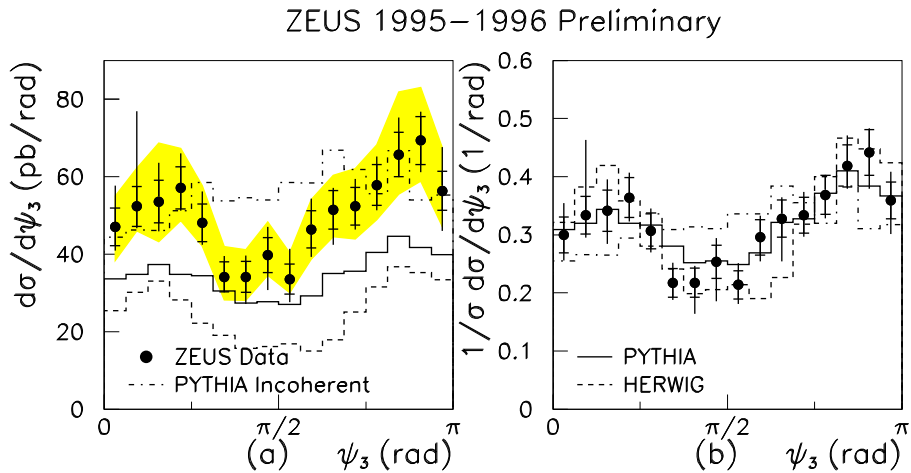}
\caption{(a) The cross section $d\sigma/d\psi_3$ and (b) the normalized 
distribution of $\psi_3$.  The error bars are as described for
Figure~\ref{fig:incjets}.  The PYTHIA and HERWIG predictions are shown by
the solid and dashed lines and the dashed-dotted line 
shows the PYTHIA prediction with colour coherence
switched off.}
\label{fig:psi_coh}
\end{figure}

The QCD phenomenon of colour coherence may be expected to favour
topologies in which the softest jet is radiated close to the plane
containing the highest energy jet and the beam.  
Figure~\ref{fig:psi_coh} shows $d\sigma/d\psi_3$ and 
$1/\sigma d\sigma/d\psi_3$ compared with the expectations from
HERWIG, default PYTHIA, and PYTHIA with coherence switched off.
Coherence suppresses large angle emissions leading to a depletion 
of the $\psi_3$ distribution near $\psi_3 = \pi / 2$.  The incoherent
prediction is relatively flat.  Coherence in the parton shower is
required by the data.  However the current measurement precision
is insufficient to discriminate between the two models, HERWIG and
PYTHIA.

\section{Summary}

ZEUS has measured hard photoproduction in a kinematic regime where
the photon's structure has not yet been well-constrained.
In both inclusive prompt-$\gamma$
production and in dijet production the GRV photon parton densities
are favoured over the GS96 parton densities.  The data reveal
fundamental phenomena of the theory of QCD.  A suppression of the
resolved component of the photon with increasing photon virtuality
has been observed and the effect of QCD dynamics on the angular
distribution of two- and three-jet events has been studied.  New
photoproduction territory will shortly become accessible to the
HERA experiments as we move into a regime of high luminosity HERA
running.


\begin{thebibliography}{99}

\bibitem{ichep_810}ZEUS Collab., J. Breitweg et al., contribution 810 to
{\em XXIX International Conference on High Energy Physics}, 
August 1998, Vancouver.

\bibitem{HERWIG}G. Marchesini et al., \Journal{\CPC}{67}{465}{1992}.

\bibitem{ichep_812}ZEUS Collab., J. Breitweg et al., contribution 812 to
{\em XXIX International Conference on High Energy Physics}, 
August 1998, Vancouver.

\bibitem{ichep_815}ZEUS Collab., J. Breitweg et al., contribution 815 to
{\em XXIX International Conference on High Energy Physics}, 
August 1998, Vancouver.

\bibitem{ichep_816}ZEUS Collab., J. Breitweg et al., contribution 816 to
{\em XXIX International Conference on High Energy Physics}, 
August 1998, Vancouver.

\bibitem{ichep_805}ZEUS Collab., J. Breitweg et al., contribution 805 to
{\em XXIX International Conference on High Energy Physics}, 
August 1998, Vancouver.

\bibitem{ichep_800}ZEUS Collab., J. Breitweg et al., contribution 800 to
{\em XXIX International Conference on High Energy Physics}, 
August 1998, Vancouver.

\bibitem{RSEP}S. D. Ellis, Z. Kunszt and D. E. Soper, 
\Journal{\PRL}{69}{3615}{1992}.

\bibitem{Mike_93}S. Catani, Yu.L. Dokshitzer, M. H. Seymour and 
B. R. Webber, \Journal{\NPB}{406}{187}{1993}.

\bibitem{Ellis_93}S. D. Ellis and D. E. Soper, 
\Journal{\PRD}{48}{3160}{1993}.

\bibitem{dijet_96}ZEUS Collab., M. Derrick et al., \Journal{\PLB}{384}{401}{1996}.

\bibitem{PYTHIA}H.-U. Bengtsson and T.~Sj{\"o}strand, \Journal{\CPC}{46}{43}{1987}.

\end{thebibliography}
\end{document}